# Web-enabled Intelligent System for Continuous Sensor Data Processing and Visualization

Felix G. Hamza-Lup
Computer Science
Georgia Southern University
Savannah GA US
fhamzalup@GeorgiaSouthern.edu

Ionut E. Iacob
Mathematical Sciences
Georgia Southern University
Statesboro, GA, USA
ieiacob@GeorgiaSouthern.edu

Sushmita Khan
Mathematical Sciences
Georgia Southern University
Statesboro GA US
sk03732@GeorgiaSouthern.edu

## ABSTRACT

A large number of sensors deployed in recent years in various setups and their data is readily available in dedicated databases or in the cloud. Of particular interest is real-time data processing and 3D visualization in web-based user interfaces that facilitate spatial information understanding and sharing, hence helping the decision making process for all the parties involved.

In this research, we provide a prototype system for near real-time, continuous X3D-based visualization of processed sensor data for two significant applications: thermal monitoring for residential/commercial buildings and nitrogen cycle monitoring in water beds for aquaponics systems. As sensors are sparsely placed, in each application, where they collect data for large periods (of up to one year), we employ a Finite Differences Method and a Neural Networks model to approximate data distribution in the entire volume.

## 1 Introduction

The widespread of sensors in the IoT context and the emergence of Web3D standards, enable big data collection and visualization, empowering Internet collaborations among participants with different backgrounds, and facilitating the decision process through spatial data and complex information understanding.

We employ the X3D ISO standard [1] and the Finite Difference Method (FDM) computational model with a Machine Learning (ML) based enhancement, to process data from a set of sensors attached at fixed locations in the monitored environment. The proposed sensor data processing allows generation of meaningful supplementary data in the entire monitored volume, to enable generation of X3D-based colormaps with variable transparency values (α- values).

The focus is on two *static* scenarios, where the location of the sensors does not change: (1) wireless sensors located in a building to assess its thermal efficiency and (2) an automated sensor system for nitrogen cycle monitoring in an aquaponics testbed. In the first scenario, the system proposed can provide valuable insights into building design, materials and construction that can lead to significant energy savings and an improved thermal comfort. In the second scenario, the system proposed will aid in the identification of nitrogen utilization efficiencies of potential fish/vegetable cropping schedules of an aquaponics food production system. A *dynamic* scenario, where the sensors are moving (e.g. vehicular ad-hoc networks - VANETs [2]) is also possible, but not investigated herein.

The Web-based 3D visualization aids at micro level (e.g. engineers, developers, architects, etc.), as well as at the macro level, decision makers (e.g. managers, project owners, investors) to propose and implement optimal solutions.

The paper is structured as follows. Section 2 provides a brief overview of the background problems in the two scenarios above and highlights some related work. In Section 3 we give a brief overview on Artificial Neural Networks models and describe the main steps of the process. Section 4 presents the sensor system and data collection process. general description of the whole process, as well as the details of the underlying model we propose. The general description of the whole framework and the underlying model we propose, and our experimental results are presented in Section 5. The conclusions and future work are presented in the last section.

## 2 Applications Proposed for Web3D Visualization

Among the most important benefits of information visualization is that it allows stakeholders visual access to huge amounts of processed or raw data in easily digestible visual representations. Moreover, Web enabled 3D data visualization enables remote collaborations and concurrent interpretations of the data among groups of people as they would be collocated. Being able to engage as a team from anywhere in the world, generates increased productivity, flexibility and significant savings.

### 2.1 Building Interior Monitoring

Autonomous wireless sensors (AWSs) are deployed in many indoor environments where they collect data for large periods of time. Since many thermally deficient western methods of construction hardly tap into the huge potential of applying energy-efficiency technology, we proposed in [3] a prototype for generating semi-transparent (α-value) X3D thermal maps to represent temperature and humidity in large commercial and residential buildings.

Although there are many commercial simulations, analysis, and visualization tools in the construction industry, most of them rely on theoretical thermal models to make decisions on the building structural design and modifications. International standards for thermal comfort for indoor air temperature and humidity [4] lack representation, measurement methods and interpretation, dealing mainly with the perceived temperature values from the building occupants. The National Institute of Building Sciences (NIBS) proposes baseline standards [5] for thermal performance of building enclosures [6], with certain levels of high performances, measurement, and verification for design and construction of enclosure assemblies.

Various methods to evaluate the building envelope are in commercial use today among which the most frequently used is the Infrared (IR) thermal imaging. It provides an accurate estimation of the temperature in 2D, as illustrated in Figure 1 (a), however, it is just a snapshot in time and does not capture the dynamic distribution of temperature and humidity inside the rooms. Using data from a sparse set of AWSs we used linear interpolation to generate additional values [7] and displayed the aggregated X3D thermal data in relationship with the building architecture, as illustrated in Figure 1 (b).

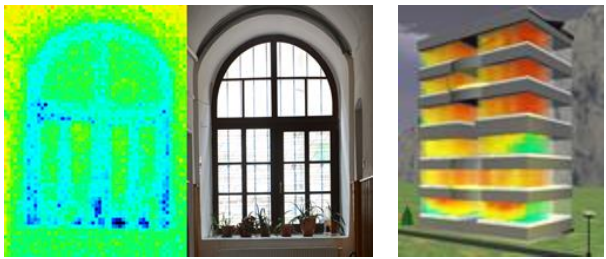

Figure 1. (a) 2D IR thermal image (b) X3D thermal maps from simulated sensor data.

Among the advantages of the proposed system are:

- Spatial Analytics: find the best location for HAVC systems, plan for a smarter building interior setup, and prepare and respond faster in emergencies, knowing the spatiality of the interior.
- Mapping and Visualization: temperature/humidity, as well as other parameters (e.g., $CO_2$ levels) monitoring help spot spatial patterns in the building and enable better decisions regarding building management.
- The web-based aspect of the application breaks down spatial barriers and facilitates collaboration among the actors involved (architects, contractors, owners etc.).
- Real-time sensor data: location monitoring of any type of sensors — will accelerate response times, optimizing safety, and improving operational awareness across all assets and activities, whether in motion or at rest.
- Potential to collect, crowdsource, store, access, and share data efficiently and securely using the XML based X3D security framework.

### 2.2 Nitrogen Compounds Monitoring

Aquaponics is a unified system that combines elements of recirculating aquaculture and hydroponics [8], greatly improving resource utilization. In aquaponics, waste water from fish tanks, enriched via feed nutrients, is used for plant growth as illustrated in Figure 2.

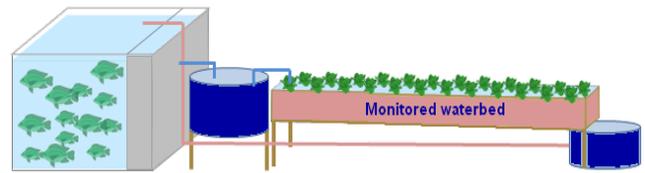

Figure 2. Aquaponics system with Nitrogen compounds monitoring – plants absorb nutrients from fish waste cleaning the water.

The interlinking of aquaculture and hydroponic procedures is considered innovative and sustainable, allowing some of the shortcomings of traditional agriculture and aquaculture systems to be addressed [9]. A well-managed aquaponics system can improve nutrient retention efficiency, reduce water usage and waste discharge (mostly lost nutrients), and improve profitability by simultaneously producing two cash crops [9, 10]. Near future food-production, will be evaluated on the ability to recover and reuse nutrients while minimizing resource loss.

Aquaponics is a sustainable alternative to traditional agriculture and aquaculture because of high assimilation rates of input nutrients, specifically nitrogen utilization efficiency [11]. Nitrogen is used in the formation of non-essential amino acids and protein for all living organisms. It is also the most expensive component in aquatic feeds (weight basis) and can account for 50–70% of operational costs in fish production [12]. Traditional aquaculture systems are relatively inefficient in terms of tissue nitrogen retention, with approximately 70% excreted into the surrounding environment in the form of ammonia [13].

Aquaponics is one of the few production techniques available that can "save" lost nitrogen from fish production processes and retain it as a secondary cash crop (e.g., green leafy vegetables).

To identify nitrogen utilization effectiveness of an aquaponics food production system, the Nitrogen compounds waterbed monitoring is required. Nitrogen monitoring requires complex sensors and frequent recalibrations. Moreover, to the best of our knowledge, there is no framework for web-based 3D visualization of Nitrogen compounds (e.g. nitrates, nitrites, ammonia) available. Nonetheless, the Nitrogen cycle is of paramount importance to every living system as it is an important part of many cells and processes such as amino acids, proteins and even our DNA.

## 3  Lightweight ML and Projected Steps

### 3.1 Succinct Background

ML is about statistics and prediction based on an existing set of data, and it can start from simple forms of Linear Regression. Data processing through classification is an essential intermediary task, and Logistic Regression (LR) can be used to classifying data into discrete results. In the LR process, a cost function has to be defined, based on which multi-class classification can occur. Once the classification is completed, the system must be able to generalize well to new examples not seen so far. For this, a regularization function is introduced to prevent the system from over-fitting training data.

The Artificial Neural Networks (ANN) models (or simply Neural Networks) have gained a very good reputation recently for their many applications in machine learning. Concisely, an ANN model aims to mimic human brain's network of neurons as a network flow where complex inputs are transformed in outputs via linear combinations of non-linear functions (called "activation functions"). ANNs have been used intensively in machine learning with applications in medical field (diagnosis of cancer), face recognition, signal classification, financial forecasting, etc. The model's strength stems from the famous "universal approximation theorem" (Cybenko [14], Hornik [15]) which essentially states that a feed forward network using a sigmoid function as an activation function can approximate any continuous function, with any precision. Hornik et al [16] improved this result using a bounded function as an activation function. Lagaris et al [17] introduced a general framework for solving initial value problems using ANN models. This new area of applications received continuous interest [18], as the approximate solutions produced by ANN models are analytical, rather than numerical.

ANNs have been traditionally used for performing machine learning classification (the famous MNIST classification problem [20] is one of the classical examples). In such traditional setup, the ANN model uses a training data set of observations whose categories (classes) are previously known to learn the parameters of the model so that classes of new observations (of categories previously unknown) are successfully predicted. The model's accuracy is typically evaluated on another set of observations of previously known classes (the validation data set). In this work we do not use ANN in the classical way for machine learning classification. We use an ANN model as an approximation model, for the best approximation mapping between a set of input values and a set of output values.

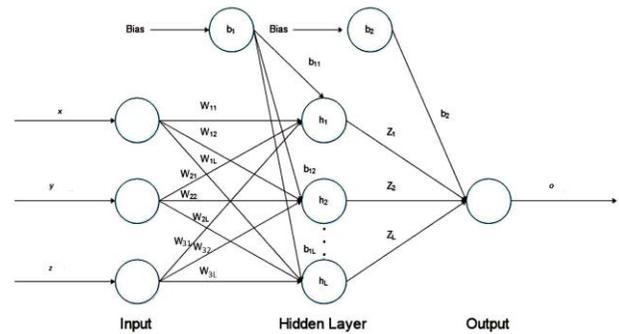

Figure 3. Artificial Neural Network (ANN) model

An Artificial Neural Network (ANN) model with one hidden layer and $L$ hidden neurons is represented in Figure 3. We use this model (with three inputs $x, y, z$; $L$ hidden nodes; and one output $N$) to approximate an unknown function $\mathbb{R}^3 \xrightarrow{u} \mathbb{R}$ as:

(1) $\quad u(x,y,z) \approx N(x,y,z) = \sum_{i=1}^{L} Z_i h_i + b_2$

where

$$h_i = \sigma(W_{1i}x + W_{2i}y + W_{3i}z + b_{1i})$$
$$\sigma(x) = \frac{1}{1+e^{-x}}$$

In the model above, the parameters $W_{ki}, b_{1i}, b_2$ (with $k = 1,2,3; i = 1 \dots L$) are learned during the neural network training process with a given data set (training set) $S \subset \mathbb{R}^4$ so that for each $(x, y, z, o) \in S$, $N(x,y,z)$ best approximates $o$ (with respect of some error measure). The universal approximation theorem guarantees that such model can approximate any continuous function $u(x,y,z)$ with arbitrary precision. However, the number $L$ of the hidden neurons for achieving certain precision can only be determined by practical trials.

### 3.2 Data Acquisition and Processing Stages

Our previous work on thermal models included an overly simplified linear interpolation model for estimating room heat distribution. While superior to standard thermal imaging, we clearly cannot capture a more sophisticated room's interior heat distribution using linear approximations. For instance, a window placed on some of the room's wall, walls shared with neighboring rooms, and exterior walls facing sun or shade would introduce significant non-linearity in the model.

In our current work, we employ the heat equation model, which governs the heat diffusion more accurately. In previous work, laminar flow models were considered for building components and small room size analysis. Chung [19] uses heat-flux sensors in conjunction with surface temperature information from thermal imaging to compute the heat-transfer coefficients at spot locations on a building facade. Our current approach (illustrated in Figure 4) is a three steps process: (i) collect real-time data from sensors, (ii) process the data, and (iii) visualize the information as a spatio-temporal matrix.

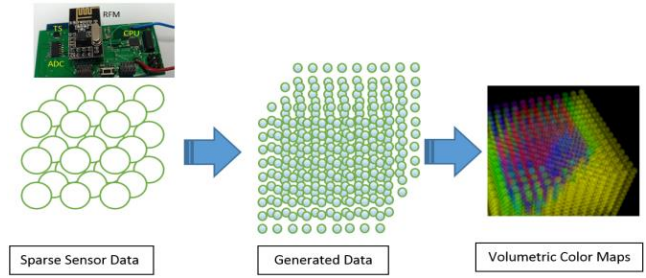

Figure 4. Steps in Data Acquisition, Processing and Display

Similar to our previous work, we collect temperature information from sensors placed on the boundaries of the enclosed volume (e.g. the 8 corners of a rectangular room, the 6 centers of each face/wall, etc.). Discrete temperature values in a few points of the considered domain (room corners, for instance) would not provide sufficient information to compute numeric solutions of the heat equation model. The complete distribution of the temperature values on the whole domain boundaries would be necessary. In the next step we perform interpolation to determine the temperature distribution on the entire room boundaries. Then, based on the heat equation model, we compute the temperature distribution at each point in the room volume. In a first computational step, we use Finite Differences (FD) techniques to compute temperature distribution at regular grid points in the 3D volume. We then apply a Neural Network approximation model to obtain an analytical formula that computes the room temperature at any point in the room (or nitrogen compounds concentration at any point in the water testbed). With this approach, we can not only refine the grid points at which the temperature/Nitrogen concentration is computed, but we can also, for instance, "zoom in" into any subdomain and analyze the temperature/Nitrogen distribution in small detail. This refinement presents some clear advantages:
1. An analytical formula is clearly a more compact form of storing the sensor information (as opposed to a large 3D array of values at each grid point). In addition, the formula can estimate temperature/Nitrogen compounds values as 3D arrays of various dimensions without significant computational effort.
2. We can use this approach to produce short-term estimates of temperature/Nitrogen compounds evolution, which, in conjunction with further sensor readings, can help detecting possible anomalies in the sensory data, such as sudden changes in the readings, which can be accounted for special events (e.g. building fire, waterbed failure, sensor failure, major intrusion event etc.).
3. The distribution of temperature/Nitrogen compounds over the entire room/waterbed volume is formatted according to the X3D specifications and, subsequently, displayed on web.

## 4 Sensor System and Data Collection

To collect the data a set of AWS is used. For thermal data collection a cost effective set of temperature/humidity sensors SHT21 [21] with potential for mobility were combined within and Arduino Uno setup. In the aquaponics setup the nitrogen in the body of liquid is more difficult to measure. An Orbisphere 315xx [22] family can be employed to compute the $N_2$ dissolved in liquid. For the Ammonia ($NH_3$) and Nitrate ($NO_3$) the ISENH3181 and the ISENO3181 [23] Ion selective electrodes can be employed, however they require manual calibration, hence not a fully automated procedure.

### 4.1 Sensor Placement and a Recursive View

In our experiments we consider that each room/volume has the shape of a cube. (We emphasize we do not sacrifice generality with this model as a more general case of a rectangular parallelepiped can be scaled to a cube.) Going inwards we further divide the cube in sub-cubes (e.g. an edge division in half would generate eight cubes of equal size in each room etc.). Going outwards, each building consists of cubic rooms and in the most simplistic scenario, the entire building is a cube (as illustrated in Figure 5).

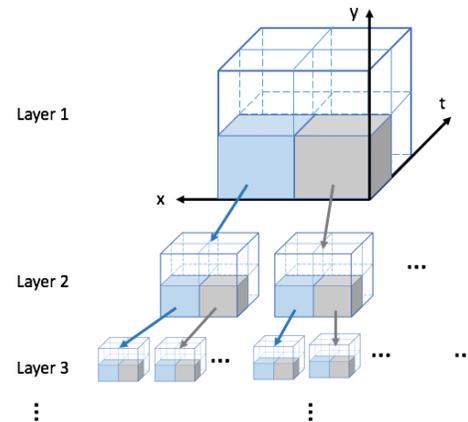

Figure 5. Recursive Geometrical Decomposition

The AWS placement strategy consists in placing an AWS in each corner of the room and six on the center of each wall with and additional sensor in the middle of the room. The AWS hybrid power system as well as the communication protocols allow data collection for extended periods – up to 1 year. A myriad of alternative sensor distribution strategies are available (e.g., based on non-laminar air/water flow) and a preliminary analysis can be done to identify best placement locations (e.g., windows, doors, HAVC location intake/output etc.), however such analysis is beyond the scope of this paper.

This recursive geometric view model comes with twofold benefits. First, it allows a recursive implementation of the model at various levels of granularity. Second, it could also be applied to non-rectangular shapes of room and/or buildings by decomposing their geometry into rectangular building blocks (hence generality).

### 4.2. Data Communication

Data collection from each set of AWS is done through a data concentrator (DC) associated with each room [4]. Data communication between the sensors and the DC is based on the idea of circular memory buffers, physically located on the end-points. These buffers are written by the microcontroller based on the data collected from each sensor. Each line in the file contains information like the sensor id and read value. The buffer implements a FIFO strategy and it can hold up to 1000 records. Consequently, by design, different query frequencies can be used by the DC. Frequencies of one query per minute were explored and proved to be suitable for medium interval data collection (up to 6 months).

We have implemented a client-server socket system that allows data delivery from the DC directly to a cloud server. The average sensor data to server, packet transmission time over a period of 24 hours is 80ms. Without replication services and within the same Metropolitan Area Network the average packet reading time from the server is 95ms (with jitter average of 30ms). The data processing delays averages is 112ms, without step 3 as described in the previous section. Therefore, by building a data-fetch buffer of 500ms we can compensate for the overall packet delay, and obtain a continuous FPS of 25+ interactive scene. The limiting factors of this experimental data is the jitter as the delay variation larger than the buffer size (i.e., 500ms) could generate data loss and inconsistencies, however, further network simulations studies in this direction are beyond the scope of the paper and will be reported elsewhere.

## 5 Predictive Model for Data Stuffing

### 5. 1 Framework

Mathematically, we consider each building room as a 3D Cartesian grid $\Omega = L \times H \times D$, where a number of temperature sensors are placed in a few grid points of $\Omega$. The temperature diffusion in each room (considered a homogeneous environment) is described by the heat flow equation:

$$\frac{\partial u}{\partial t} - \alpha \left( \frac{\partial^2 u}{\partial x^2} + \frac{\partial^2 u}{\partial y^2} + \frac{\partial^2 u}{\partial z^2} \right) = 0$$

Where, u(x,y,z,t) is the temperature function at each spatial coordinate (x,y,z) and time t. For simplicity (but without sacrificing much of the purpose of our analysis), we carry our analysis for the moments of time when the steady state (or close) has been reached. For this case scenario, the equation above becomes the Laplace equation:

$$\frac{\partial^2 u}{\partial x^2} + \frac{\partial^2 u}{\partial y^2} + \frac{\partial^2 u}{\partial z^2} = 0$$

On specific domains and given appropriate boundary conditions the Laplace equation above has an analytical solution. Otherwise, numerical solutions are the only options if complete boundary information is provided. Our first challenge consists of computing the boundary values to completely solve the problem. The only information we start with consists in a finite number of discrete temperature values, such as eight temperature readings from all the corners of a rectangular room and/or room walls centers (Figure 5, left). Consequently, before proceeding to numerically solving the diffusion equation we must interpolate the given values and compute values on the given domain boundaries (all faces of a rectangular prism for a rectangular room). For this purpose, we used two methods: (i) a linear 2D interpolation (on each face of the cube) and (ii) a 2D heat diffusion model for each face of the cube (domain). With either method, we obtain a heat diffusion model described by a Laplace equation with Dirichlet boundary conditions (where $\partial\Omega$ denotes the boundary of the domain $\Omega$):

(2) $$\frac{\partial^2 u}{\partial x^2} + \frac{\partial^2 u}{\partial y^2} + \frac{\partial^2 u}{\partial z^2} = 0 \qquad on\ \Omega$$
$$u(x, y, z) = f(x, y, z) \qquad on\ \partial\Omega$$

We proceed to solve this equation numerically, using the classic finite differences (FD) method. However, FD is known to be computationally expensive. Producing numerical results with finer granularity (on the whole domain or subsets of the domain) would require a new FD solution. We alleviate this issue by computing only once the FD numerical solution, for an intermediate-size domain grid (hence not very expensive to compute). We then refine the solution using a Neural Network model, which will produce an approximation function:

$$\tilde{u}(x, y, z) \approx u(x, y, z) \qquad on\ \Omega$$

Not only the above function can quickly compute values at *any* domain point, but also it can be considered a canonical representation of *all* values in the domain. These are certainly highly desired characteristics for a real-time process.

### 5.2 Full Volume Data Generation

We simulated our computational framework (Figure 4) using the Finite Differences (FD) and Artificial Neural Networks (ANN) models on the $\Omega = (0,1)^3$ domain (a cube). The initial sensor data was considered for two case scenarios:

(1) All corners of the cube (8 data points):
$$S_1 = \{u(i, j, k) \mid i, j, k = 0,1\}$$
(2) All cube's corners, plus centers of each face (8+6 = 14 data points):
$$S_2 = S_1 \cup \{u(0.5, 0.5, 0), u(0.5, 0, 0.5), u(0, 0.5, 0.5),$$
$$u(0.5, 0.5, 1), u(0.5, 1, 0.5), u(1, 0.5, 0.5, 0.5)\}$$

Table 1. Sensor data values for experimental simulations

| $u$ value | Location |
|---|---|
| 19 | front-bottom-left |
| 20 | front-bottom-right |
| 26 | front-top-left |
| 27 | front-top-right |
| 20 | back-bottom-left |
| 21 | back-bottom-right |
| 25 | back-top-left |
| 26 | back-top-right |

Table 1 shows a snapshot of sensor reading values in $S_1$ used in our experimental simulation. The full volume data $\tilde{u}(x,y,z)$ was generated as follows:
1. Boundary values for the model in Equation (2) were computed from $S_1$ (or $S_2$) data by linear interpolation (or using the FD method for a 2D version of the model in Equation (2)) on a $8 \times 8$ grid for each cube's face.
2. Using the FD method for Equation (2), all values of the regular $6 \times 6 \times 6$ interior grid were computed, and hence the full $8 \times 8 \times 8$ grid data was fully available:
$$S = \left\{ \tilde{u}\left(\frac{i}{7}, \frac{j}{7}, \frac{k}{7}\right) \mid i,j,k = 0..7 \right\}$$
The result of this step (the set of grid points $S$) is presented in Figure 6 (left).
3. Next, we used an ANN model as in Equation (1) with the training set $S$ from step 2 above to construct the continuous, analytical function $\tilde{u}(x,y,z)$ that fully approximates $u(x,y,z)$ on the whole domain $\Omega = (0,1)^3$. Numerical computation results on a refined $16 \times 16 \times 16$ grid are shown in Figure 6 (right).

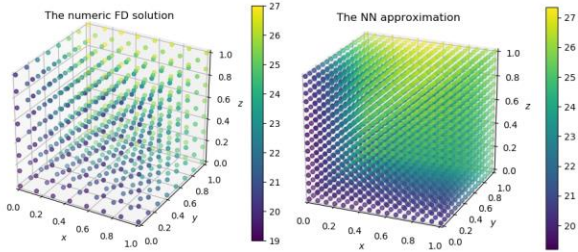

Figure 6. Refinement from FD using ANN

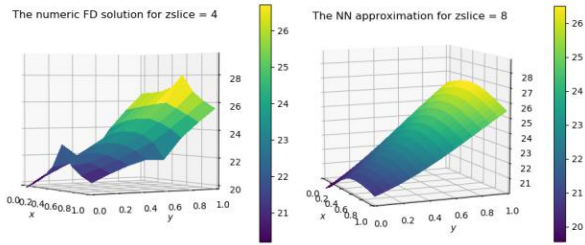

Figure 7. Smoothing FD approximation using ANN

Training an ANN model as in step 3 above is, in general, computationally expensive and time consuming (depending on the computational resources being used and/or the size of the grid in step 2). While it brings the clear benefits of constructing an analytical approximation that can produce values in all points of the domain (Figure 7 shows how the ANN model smoothers the numerical solution obtained using FD methods), it may be a burden in real-time monitoring systems. In practice, when the ANN training time far exceeds the period of sensor data delivery, this step does not need to be performed for each sensor data reading. For instance, on a regular Intel Core i5-62000U CPU @ 2.30GHz 8GB RAM computer, we performed steps 1 and 2 in 112ms while step 3 was performed in 64s. However, in practical applications, step 3 needs not be performed for every reading as high-granularity visualization is not continuously needed. Instead, step 3 can be performed, for instance, while zooming in on stationary or historical data (playback mode) when its time performance is acceptable even on average/low power computational platforms.

### 5.3 X3D-based Visualization
The full volume data $\tilde{u}(x,y,z)$ (Figure 6, left) is subsequently converted to X3D format for visualization. For the X3D representation, we used the Box primitive and semitransparency values that allow in volume visualization as illustrated in Figure 8.

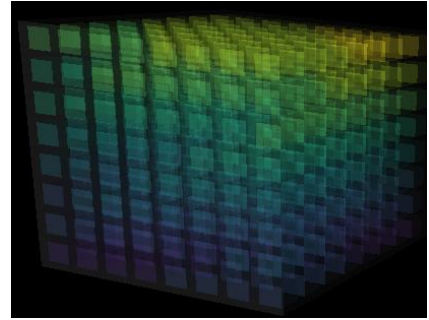

Figure 8. X3D based visualization of diffused temperature maps.

For the aquaponics waterbed we implemented a basic setup using predefined values as illustrated in Figure 9.

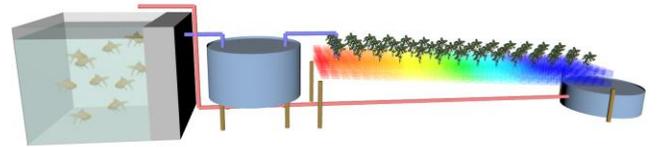

Figure 9. X3D based visualization of aquaponics waterbed Nitrogen compounds.

### Conclusion
Autonomous wireless sensors represent the senses of the environment by enabling data collection from a variety of real-world applications. The IoT infrastructure enables a structural approach to data collection, which, used in conjunction with ML may allow fast real-time data processing.

Web-based 3D information visualization allows visual access to huge amounts of data in easily digestible visual representation, as well as engagement as a team from anywhere in the world in problem solving and decision-making. We exemplify with two important real-world applications: building thermal efficiency assessment and nitrogen compounds concentration quantification in aquaponics waterbeds.

A possible extension of our framework we would like to investigate in the future is using the ANN model (or a some other machine learning technique) to learn from the recent inside temperature history and short term outside temperature forecast to produce energy efficient recommendations for thermostat temperature settings and to allow interactive Web-based 3D visualization of processed data.

For the applications in aquaponics, being able to track the Nitrogen compounds in the water testbed is of paramount importance for the systems equilibrium and efficiency. Such miniature aquafarms have the potential to be deployed in various regions of the world as sustainable food producers, and help alleviate world hunger problems.

## ACKNOWLEDGMENTS

This work was also facilitated by the Erasmus+ program for international cooperation between Georgia Southern University, US and Transilvania University of Brasov, Romania.

## REFERENCES

[1] Polys, N. F.; Brutzman, D.; Steed, A.; Behr, J. Future Standards for Immersive VR: Report on the IEEE Virtual Reality 2007 Workshop. IEEE Computer Graphics and Applications, 28 (2):94-99, 2008, Extensible 3D (X3D), Web3D Consortium. Available online: www.web3d.org (accessed on 09. 09. 2018).

[2] A. A. Rashid, R. N. Mahapatra, and F. G. Hamza-Lup (2019) "Adaptive Group-based Zero Knowledge Proof-Authentication Protocol (AGZKP-AP) in Vehicular Ad Hoc Networks," IEEE Transactions of Intelligent Transportation Systems.

[3] Hamza-Lup, F. G.; Borza, P. N.; Dragut, D.; Maghiar, M. X3D Sensor-based Thermal Maps for Residential and Commercial Buildings, Web3D, 2015, Heraklion, Crete, Greece, June 18-21.

[4] Nicol, F., Humphreys, M., Sykes, O., and Roaf, S. (1995). Standards for thermal comfort. TJ Press Ltd., UK

[5] NIBS (2019). National Performance Based Design Guide. Building Envelope Design Guide. Available online at: http://www.wbdg.org/guides-specifications/building-envelope-design-guide. Retrieved Mar.6, 2019.

[6] Ashrae, 2019. Available online at: https://www.ashrae.org/technical-resources/standards-and-guidelines.

[7] Hamza-Lup, F. G.; Maghiar, M. Web3D Graphics enabled through Sensor Networks for Cost-Effective Assessment and Management of Energy Efficiency in Buildings, Graphical Models Journal, Elsevier, 2016.

[8] USDA 2019, https://www.nal.usda.gov/afsic/aquaponics

[9] Tyson, R. V., Treadwell, D. D., & Simonne, E. H. (2011). Opportunities and Challenges to Sustainability in Aquaponic Systems, HortTechnology hortte, 21(1), 6-13. Retrieved Mar 13, 2019, from https://journals.ashs.org/view/journals/horttech/21/1/article-p6.xml.

[10] Klinger, Dane, Naylor, Rosamond (2012) Searching for Solutions in Aquaculture: Charting a Sustainable Course. Annual Review of Environment and Resources. https://doi.org/10.1146/annurev-environ-021111-161531.

[11] C. Somerville, M. Cohen, E. Pantanella, A. Stankus, A. Lovatelli Small-scale aquaponic food production: integrated fish and plant farming U. FAO (Ed.), FAO Fisheries and Aquaculture Technical Paper (2014), pp. 1-262 Rome, Italy

[12] Valente, L.M.P., Linares, F., Villanueva, J.L.R., Silva, J.M.G., Espe, M., Escorcio, C., Pires, M.A., Saavedra, M.J., Borges, P., Medale, F., Alvarez-Blazquez, B., Peleteiro, J.B., (2011). Dietary protein source or energy levels have no major impact on growth performance, nutrient utilisation or flesh fatty acids composition of market sized Senegalese sole. Aquaculture 318 (1–2): 128–137.

[13] Hargreaves J. A., 1998 Nitrogen biogeochemistry of aquaculture ponds. Aquaculture 166:181-212.

[14] Cybenko, G. (1989) "Approximations by superpositions of sigmoidal functions", Mathematics of Control, Signals, and Systems, 2(4), 303–314.

[15] Kurt Hornik (1991) "Approximation Capabilities of Multilayer Feedforward Networks", Neural Networks, 4(2), 251–257.

[16] Hornik, K., Stinchcombe, M., White, H., (1989) "Multilayer feedforward networks are universal approximators", Neural Networks, 2(5), 359-366.

[17] I. E. Lagaris, A. Likas and D. I. Fotiadis, "Artificial neural networks for solving ordinary and partial differential equations," in IEEE Transactions on Neural Networks, vol. 9, no. 5, pp. 987-1000, Sept. 1998.

[18] Han, Jiequn, Jentzen, Arnulf, et al. "Overcoming the curse of dimensionality: Solving high-dimensional partial differential equations using deep learning". arXiv preprint, arXiv:1707.02568, 2017.

[19] Chung, D. (2015). Historic Building Façades: Simulation, Testing and Verification for Improved Energy Modeling. Journal of the National Institute of Building Sciences, Feb., (3), No. 1, 16-21.

[20] Ciresan, Dan; Ueli Meier; Jürgen Schmidhuber (2012). Multi-column deep neural networks for image classification. 2012 IEEE Conference on Computer Vision and Pattern Recognition. pp. 3642–3649.

[21] SENSIRION (2019) "Sensirion Relative-Humidity Sensor". Specifications online at: http://www.sensirion.com/. Accessed March 10, 2019

[22] Orbisphere (2019) https://www.hach.com/nitrogen-sensors/orbisphere-315xx-nitrogen-sensors/family?productCategoryId=35547372750

[23] ISENO (2019) https://www.hach.com/hq440d-water-quality-laboratory-ise-ammonia-nh-and-nitrate-no-sup-sup-ion-meter-package-with-isenh3181-iseno3181-ion-selective-electrodes/product?id=7640592116&callback=qs